\documentclass[aps,twocolumn,amsmath,amssymb, superscriptaddress, nofootinbib]{revtex4}
\usepackage[pdftex]{graphicx}
 \usepackage{amsmath}
\usepackage[T1]{fontenc}
\usepackage{amssymb}
\usepackage{amsfonts}
\usepackage{bm}
 \usepackage{amsmath} 
  \usepackage{flushend}
\usepackage{color}
\usepackage{lipsum}
\usepackage{amsfonts} 
\usepackage{amssymb, mathrsfs}
\usepackage{braket}
\usepackage{graphicx} 
\usepackage{subfigure}
\usepackage{bbm}
\def\beq{\begin{equation}}
\def\eeq{\end{equation}}
\def\bsp{\begin{split}}
\def\esp{\end{split}}
\def\bea{\begin{eqnarray}}
\def\eea{\end{eqnarray}}
\def\ba{\begin{array}}
\def\ea{\end{array}}

\def\dg{\dagger}

\def\l.{\left.}
\def\r.{\right.}

\def\ra{\rangle}
\def\la{\langle}

\def\bo{{\vec k}}

\begin{document}

\date{\today}
\title{Magnonic Floquet Hofstadter Butterfly}
\author{S. A. Owerre}
\affiliation{Perimeter Institute for Theoretical Physics, 31 Caroline St. N., Waterloo, Ontario N2L 2Y5, Canada.}

\begin{abstract}
We introduce the magnonic  Floquet Hofstadter butterfly in the two-dimensional insulating honeycomb ferromagnet.  We show that when the insulating  honeycomb ferromagnet is irradiated by an oscillating space- and time-dependent electric field,  the hopping magnetic dipole moment (i.e. magnon quasiparticles) accumulate the Aharonov-Casher phase.  In the case of only space-dependent electric field, we realize the magnonic Hofstadter spectrum with similar fractal structure as graphene subject to a perpendicular magnetic field, but with  no spin degeneracy due to broken time-reversal symmetry by the ferromagnetic order. In addition, the magnonic Dirac points and Landau levels occur at finite energy as expected in a bosonic system. Remarkably, this discrepancy does not affect the topological invariant of the system.  Consequently, the magnonic Chern number assumes odd values and the magnon Hall conductance gets quantized by odd integers. In the case of both space- and time-dependent electric field, the theoretical framework is studied by the Floquet formalism. We show that the magnonic  Floquet Hofstadter spectrum emerges entirely from the oscillating space- and time-dependent electric field, which is in stark contrast to electronic Floquet Hofstadter spectrum, where irradiation by circularly polarized light and a perpendicular magnetic field are applied independently.    We study the deformation of the fractal structure at different laser frequencies and amplitudes, and analyze the  topological phase transitions associated with gap openings in the magnonic Floquet Hofstadter butterfly. 
\end{abstract}
\maketitle

\section{Introduction}

A plethora of interesting phenomena is manifested when a two-dimensional (2D) electronic gas system is subject to a  perpendicular  magnetic field. A well-known phenomenon  is  the integer quantum Hall effect \cite{kli,thou,lau1,nov, tsu,thou1, nov1, gus}. Theoretically, the integer quantum Hall effect is understood as a consequence of the Landau level  quantization of classical cyclotron orbits of an electron in the presence of a perpendicular magnetic field.  For relativistic electron in graphene,  the zeroth Landau level has a quantum anomaly or reduced degeneracy as it  is shared between electrons and holes. This results in an unconventional integer quantum Hall effect \cite{nov, gus}. The Hofstadter butterfly \cite{hof,ram} also emerges  when a perpendicular magnetic field is applied to a periodic electronic crystal lattice, and has been experimentally realized in some systems \cite{pono, dean, hunt}.    Beside the effects of a perpendicular magnetic field,  periodically driven electronic systems have garnered considerable attention as a mechanism for engineering Floquet topological bands \cite{foot3, foot4, fot1, jot, eck1, foot5, eck2, eck3, tp6, tp7, tp8, tp6a} in topologically trivial metallic systems. Recently, a handful of studies  have examined  the combined effects of a periodic drive and a perpendicular magnetic field in the context of the electronic Floquet Hofstadter butterfly \cite{wac, wac1,wac2, lab, lab1,lab2}.

Remarkably, the recent experimental realization of  2D honeycomb ferromagnet in CrI$_3$ and other materials \cite{Huang, Gong} has provided  a great  possibility by which the magnonic analog of graphene can be realized. In fact, a recent inelastic neutron scattering experiment has confirmed the existence of topological magnons in CrI$_3$ \cite{chen}, which follows from a previous theoretical proposal \cite{sol}. There is no doubt that magnon  holds the future of dissipationless spin transport in insulating magnets \cite{tak4}, due to its  charge-neutrality and spin precession. This active field of study is currently known as magnon spintronics \cite{chu, benja}. It involves the transport of spin and  magnetic dipole moment of  magnon  in analogy  to the transport of spin and charge of an electron in conducting materials. Moreover, the topological aspects of magnons in insulating magnets  are currently  an active research field \cite{chis, bao, fei,mok,owerre, lau, mccla,jos, ruck, xian, boy, skim, per}.

Magnon quasiparticles in insulating magnets are simply hopping  magnetic dipole moments on the lattice. Hence,  in the presence of an electromagnetic field, they accumulate the Aharonov-Casher phase \cite{aha,loss, ahaz, ahat, mei, bak, su}, analogous to the Peierls phase which  charged particles accumulate in the presence of a perpendicular magnetic field.  Based on this formalism,  K. Nakata et al. \cite{mei} have studied the magnonic analog of the integer quantum Hall effect in the 2D insulating square-lattice magnet, subject to a time-independent spatially-varying electric field gradient, which plays the same role as a perpendicular magnetic field in electronic charged system.  In this non-relativistic magnonic system, the lowest nearly flat band gives rise to a low-temperature quantized magnonic Hall conductance. Interesting features are expected to emerge in the 2D insulating honeycomb magnets with relativistic Dirac magnon. However, the quantization rule for the relativistic Dirac magnon is vaguely known.

 In this paper, we introduce another concept --- the magnonic Floquet Hofstadter butterfly.    We consider a  2D insulating honeycomb ferromagnet irradiated by an oscillating space- and time-dependent electric field. 
 
This paper is divided  into two parts.  In the first part, we study the effects of the space-dependent part of the electric field on the 2D insulating honeycomb ferromagnet. We realize the magnonic Hofstadter spectrum with similar fractal structure as the electronic honeycomb lattice \cite{hof,ram, hat, yas, gold}. The major difference in the current insulating system is that the Dirac points occur at finite energy and the spin degeneracy is lifted due to broken time-reversal symmetry by the magnetic order. Therefore, the Landau levels have only valley degeneracy. Despite these discrepancies, we show that the topological invariant of the 2D insulating honeycomb ferromagnet is unchanged from that of electronic honeycomb lattice \cite{nov, gus, hat, yas, gold}. 
 
 In the second part, we study the interaction of light with magnon-Bloch states on the  2D insulating honeycomb ferromagnet using the Floquet formalism. We show that the magnonic Floquet Hofstadter butterfly can be generated entirely from the oscillating space- and time-dependent electric field. This formalism is indeed different from the electronic Floquet Hofstadter butterfly, which is generated by independently applying circularly polarized light and a perpendicular magnetic field \cite{wac, wac1,wac2}.  We further investigate the deformation of the  fractal structure by radiation and  the topological phase transition  of the underlying magnonic multiband Hofstadter butterfly.

 \section{Model} 
 \subsection{Heisenberg spin model}
 We consider the Heisenberg spin Hamiltonian for 2D insulating honeycomb ferromagnet in the presence of a Zeeman magnetic field
 \begin{align}
\mathcal H&=-J\sum_{ \la i,j\ra }{\vec S}_{i}\cdot{\vec S}_{j}-g\mu_B \vec B\cdot\sum_{i}  \vec{S}_{i},
\label{spinh}
\end{align}
where $J>0$ is the Heisenberg ferromagnetic coupling between the nearest-neighbour (NN) spins, $\vec B = B\hat z$ is the Zeeman magnetic field applied along the $z$-direction, $g$ is the Land\'e g-factor, and $\mu_B$ is the Bohr magneton. Note that the magnetic field is only required to polarize the ferromagnetic spins along the $z$-axis. Thus, we shall set it to zero in the subsequent sections. 
\begin{figure}
\centering
\includegraphics[width=.9\linewidth]{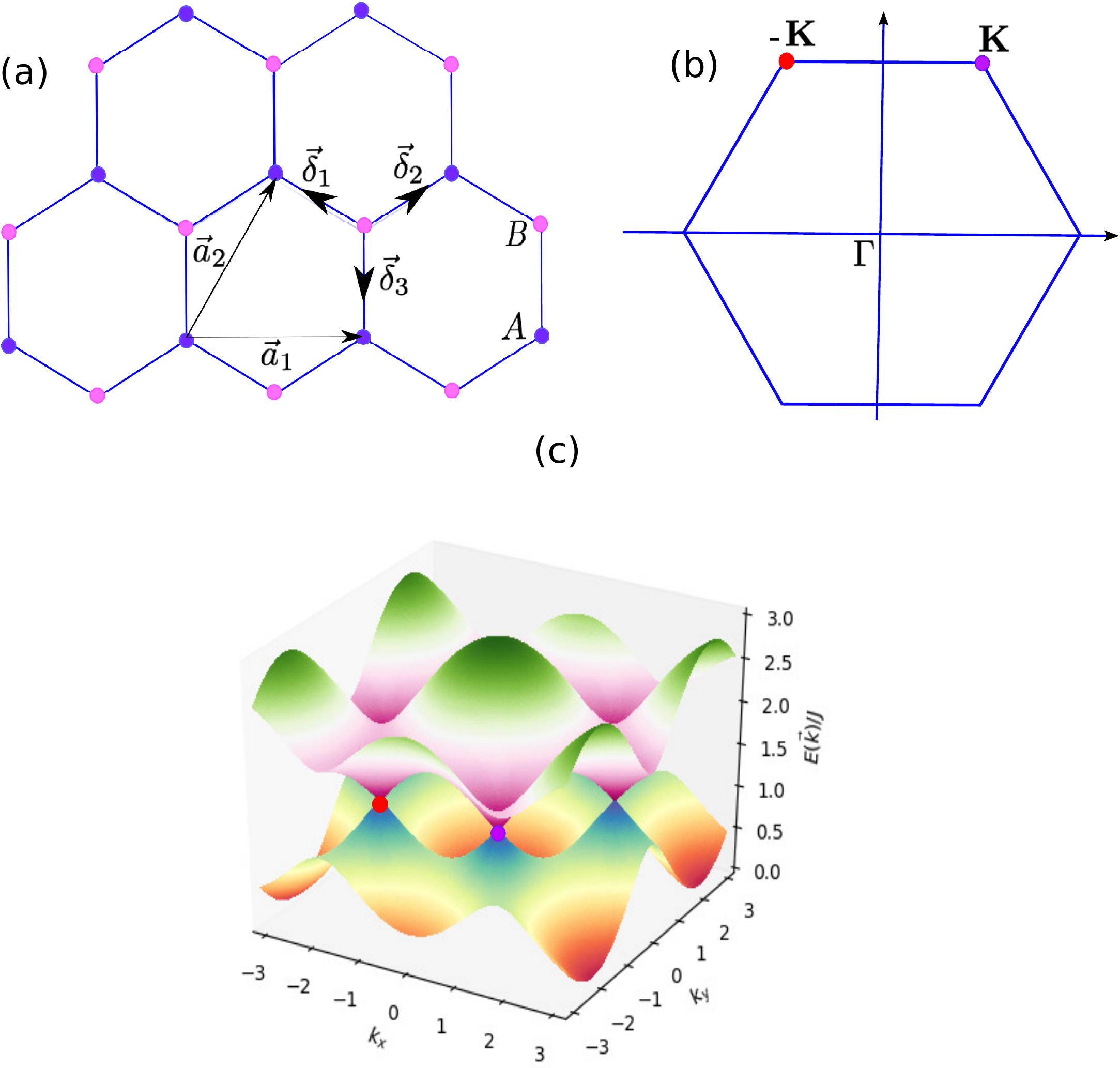}
\caption{Color online. (a) Schematic of the honeycomb lattice. The primitive lattice vectors   are   $\vec a_{1}=a\sqrt{3}\hat x$, $\vec a_{2}=a(\sqrt{3}\hat x/2 + 3\hat y/2)$, and the nearest-neighbour vectors are $\vec{\delta}_{1,2}=a(\mp\sqrt{3}\hat x/2 + \hat y/2)$, $\vec{\delta_3}=-a\hat y$.  (b) The Brillouin zone (BZ) of the honeycomb lattice. The two independent Dirac points are indicated by red and pink dots.  (c). The Dirac magnon bands of 2D honeycomb ferromagnet. }
\label{lattice}
\end{figure}

 \subsection{Bosonic tight-binding model}

 We are interested in the low-energy magnetic excitations of the spin Hamiltonian   in Eq.~\ref{spinh}.  In the low-temperature regime the magnetic excitations can be described  by  the Holstein Primakoff (HP)  transformation \cite{hospri}. Consequently, the spin Hamiltonian maps to a bosonic tight binding hopping model 
\begin{align}
 \mathcal H&=J_0S\sum_{m,n} \big(a_{m,n}^\dagger a_{m,n}+b_{m,n}^\dagger b_{m,n}\big)\nonumber\\&-JS\sum_{ m,n}\Big[a_{m,n}^\dagger\big(b_{m,n}+b_{m-1,n}+b_{m,n-1}\big)+ \text{H.c.}\Big],
\label{tight1}
\end{align}
where $J_0 = 3J + g\mu_B B/S$.  The sum is taken over all the unit cell positions. The position of an arbitrary unit cell is given by $\vec R_{mn} = m\vec{a}_1+n\vec{a}_2$, where $(m,n)$ are the unit cell indexes.   The chosen lattice vectors  of the honeycomb lattice $\vec a_{1,2}$  are depicted in Fig.~\ref{lattice}(a). Here, $a_{m,n}^\dagger(b_{m,n})$ are the creation and annihilation operators in sublattice $A(B)$. They satisfy the bosonic commutation relations:
\begin{align}
\big[c_{m,n},c_{m^\prime,n^\prime}^\dg\big] = \delta_{m,m^\prime}\delta_{n,n^\prime},
\end{align}
where $c_{m,n} = a_{m,n}, b_{m,n}$. Note that operators on different sublattice commute as usual.  The Hamiltonian in Fourier space is given by  
\begin{align}
\mathcal H= \sum_{\vec k}\psi_{\vec k}^\dg\cdot\mathcal H({\vec k})\cdot\psi_{{\vec k}},
\end{align}
 with   
\begin{align}
\mathcal H(\bo)&=J_0S~\mathbb{I}_{2\times 2}-JS 
\begin{pmatrix}
0 &f(\vec k)\\
f^*(\vec k)&0
\end{pmatrix},
\label{honn}
\end{align}
where $\mathbb{I}_{2\times 2}$ is an identity matrix, $\psi_{\vec k}^\dg = \big(a_{\vec k}^\dg,b_{\vec k}^\dg\big)$, and   $f(\vec k)=1+ e^{ik_1}+e^{ik_2}$ with $k_i= \vec k\cdot \vec {a}_i$. Diagonalization of the Hamiltonian leads to the magnon energy bands
\begin{align}
E_\pm(\vec k) = J_0S \pm JS|f(\vec k)|.
\end{align}
Throughout the analysis in this paper we will set  the Zeeman magnetic field to zero $B=0$ and consider spin $S=1/2$. The two magnon branches touch at the Dirac points $\pm{\bf K}$ as depicted in Fig.~\ref{lattice}(b) and Fig.~\ref{lattice}(c) \cite{mag,sol}. Due to the bosonic nature of magnon, the Dirac points occur at nonzero energy $E_{D}=J_0/2$.

\section{Time-independent magnonic Hofstadter spectrum}

\subsection{Time-independent Aharanov Casher phase}
We will first present the analysis for the time-independent magnonic Hofstadter spectrum in the 2D insulating honeycomb ferromagnet. The main objective here is to get familiar with the theoretical formalism that will be used in the subsequent sections.  We consider then the effects of a spatially-varying electric field ${\vec E}(\vec r)$,  propagating perpendicular to the insulating 2D honeycomb ferromagnet. As magnons are simply  magnetic dipole moment $g\mu_B\hat z$, hopping on the lattice, they will accumulate  the Aharanov Casher phase \cite{aha,loss, ahaz, ahat, mei, bak} 
\begin{align}
\theta_{mn} = \mu_m\int_{\vec R_{mn}}^{\vec R_{m n}^{\prime}} \mathscr {\vec E}(\vec r)\cdot d{\vec l},
\label{acp1}
\end{align}
where  $\mu_m = g\mu_B/\hbar c^2$. Here $\hbar = h/2\pi$ and $c$ are  the reduced Planck's constant and the speed of light respectively. We have used the notation $\mathscr {\vec E}(\vec r)={\vec E}(\vec r) \times\hat z$ for brevity.  The electric field  can be generated from an electromagnetic harmonic scalar potential \cite{mei} or an elastic gauge field \cite{yago}.   Note that the Aharanov Casher phase is dual to the Peierls phase accumulated by an electron hopping  in the background of a perpendicular magnetic field.

We consider the electric field $\vec E(\vec r) = \lambda (0, -y, 0)$, where  $\lambda$ is the linear charge density. This corresponds to the Landau gauge  $\mathscr {\vec E}(\vec r) = \lambda (-y, 0, 0)$, with $\vec \nabla\times \mathscr {\vec E}(\vec r) = \lambda\hat z$.  We see that  $\lambda$ plays a similar  role to a perpendicular magnetic field in electronic charged systems. In this gauge, the phase appears only for magnon hopping along the $x$ direction through the Peierls substitution $\vec k \to \vec k - \mu_m\lambda y\hat x$.  We define the magnonic analogs of flux quantum and flux per unit cell as $\Theta_0 = hc^2/g\mu_B$ and $\Theta= 3\sqrt{3}a^2\lambda/2$ respectively. Thus, the ratio $\Theta/\Theta_0$  is the dimensionless quantity for our model. The resulting bosonic tight-binding model in the presence of the magnon phase is given by
\begin{align}
\mathcal H&=J_0S\sum_{mn} \big(a_{m,n}^\dagger a_{m,n}+b_{m,n}^\dagger b_{m,n}\big)\nonumber\\&-JS\sum_{ mn}\Big[a_{m,n}^\dagger\big(e^{in\pi \frac{\Theta}{\Theta_0}}b_{m,n}+e^{-in\pi \frac{\Theta}{\Theta_0}}b_{m-1,n}\nonumber\\&+b_{m,n-1}\big)+ \text{H.c.}\Big].
\label{phase}
\end{align} 
\begin{figure}
\centering
\includegraphics[width=.95\linewidth]{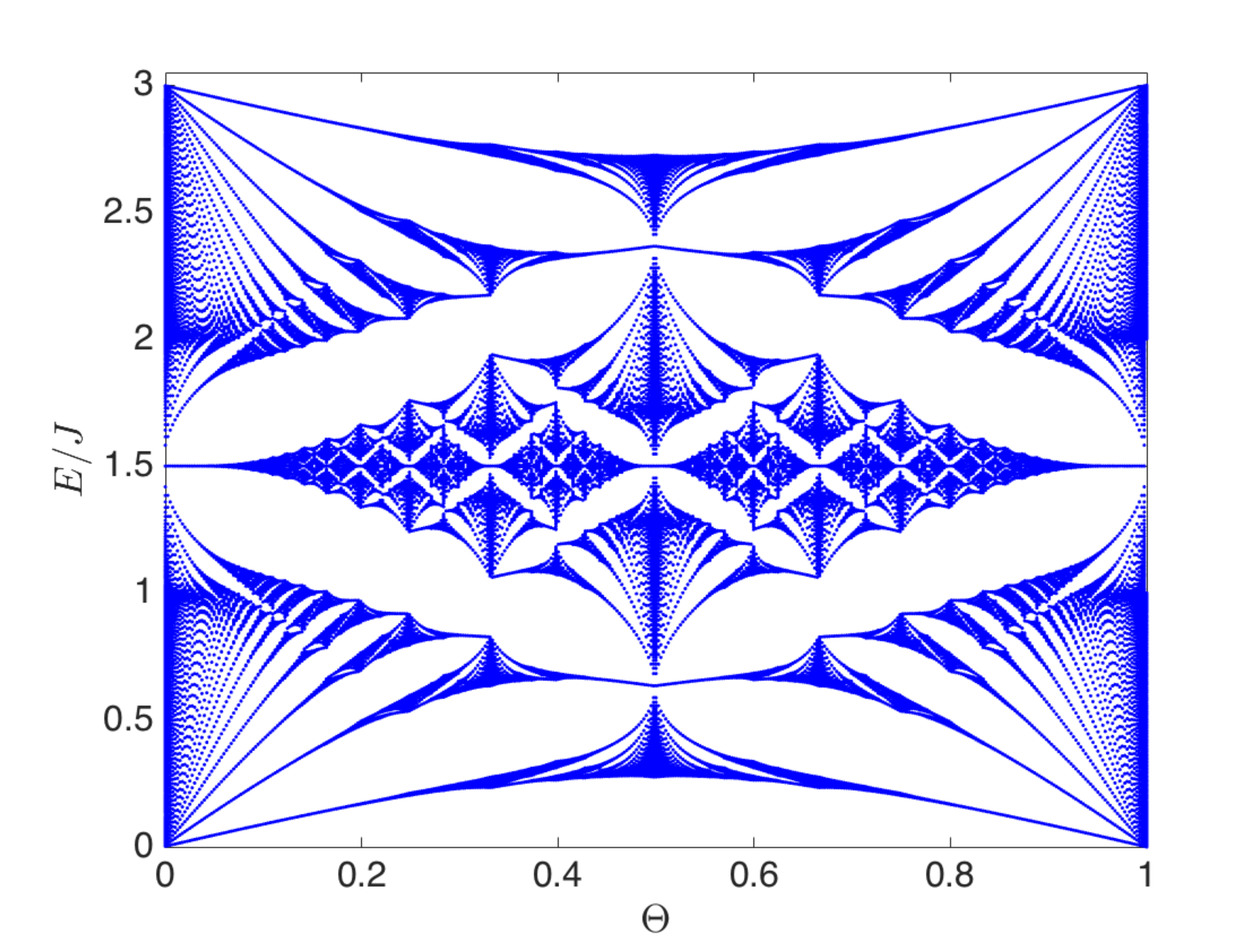}
\caption{Color online. Magnonic Hofstadter spectrum in the honeycomb ferromagnet under the influence of a time-independent  spatially-varying electric field.}
\label{HCB_Butterfly}
\end{figure}

 \subsection{Hamiltonian diagonalization}
The Schr\"ondinger wave solutions corresponding to the  Hamiltonian in Eq.~\ref{phase} can be written as \begin{align}
\ket{k} = \sum_{mn}e^{\vec k\cdot \vec R_{mn}}\big( \alpha_n a_{m,n}^\dg + \beta_n b_{m,n}^\dg\big)\ket{0},
\end{align}
where $\alpha_n $ and $ \beta_n$ are complex amplitudes, and $\ket{0}$ is the magnon vacuum state. The corresponding eigenvalue equation is given by
\begin{align}
&E\alpha_n = J_0S\alpha_n -JS\Big[\lbrace 2e^{i\frac{k_1}{2}}\cos( n\pi \frac{\Theta}{\Theta_0}-\frac{k_1}{2})\rbrace \beta_n\nonumber\\& + e^{ik_2}\beta_{n-1}\Big],\\&
E\beta_n = J_0S\beta_n -JS\Big[\lbrace 2e^{-i\frac{k_1}{2}}\cos( n\pi \frac{\Theta}{\Theta_0}-\frac{k_1}{2})\rbrace\alpha_n\nonumber\\& + e^{-ik_2}\alpha_{n+1}\Big],
\end{align}
where $\alpha_{n+q}=\alpha_{1}$ and $\beta_{n+q}=\beta_{1}$.  We consider $\Theta = p/q$, in units of $\Theta_0$, where $p$ and  $q$ are relative primes. Then the Hamiltonian  we need to diagonalize is of the size $2q\times 2q$. It can be written as \begin{align}
 \mathcal H = \int_0^{2\pi}\frac{k_1}{2\pi}\int_{0}^{2\pi/q}\frac{k_2}{2\pi/q}~\psi^\dg(\vec k)\cdot \mathcal H(\vec k)\cdot\psi(\vec k),
 \end{align}
where $\psi(\vec k)$ is a $2q$ column vector and the integration is performed over the magnetic BZ.
\begin{align}
&\mathcal H(\vec k) = J_0S~\mathbb{I}_{2q\times 2q} - JS
\begin{pmatrix}
  0 &   F({\vec k})\\
 F^\dg({\vec k}) &  0 
 \end{pmatrix},
 \label{free}
\end{align}
where $  F({\vec k})$ is  a $q\times q$ matrix  given by 
\begin{align}
& F({\vec k}) =  e^{i\frac{k_1}{2}} U+  V,
\end{align}
where
\begin{align}
V = \begin{pmatrix}
 0& 0& 0 \cdots & 0 &e^{iqk_2}\\ 
 1& 0& 0\cdots & 0 &0\\
 0& 1& 0 \cdots & 0 &0\\
 \vdots&\vdots&\vdots&\vdots&\vdots\\
  0& 0& 0 \cdots & 0 &0\\
  0& 0& 0 \cdots & 1 &0
 \end{pmatrix},
\end{align}
\begin{align}
U = \text{diag}\big( \xi_1,  \xi_2,  \xi_3, \cdots,  \xi_{q-1},  \xi_q\big),
\end{align}
with $\xi_{r} = 2\cos\big(r\pi\frac{\Theta}{\Theta_0}-\frac{k_1}{2}\big)$.

 In Fig.~\ref{HCB_Butterfly} we have displayed  the magnonic analog of the so-called Hofstadter  butterfly \cite{hof} for the 2D insulating honeycomb ferromagnet. The structure of the Hofstadter  spectrum mimics its electronic counterpart \cite{ram,yas, gold,hat}.  It possesses a  reflection symmetry about  $\Theta=1/2$ and another reflection symmetry about the Dirac magnon energy $E_D$.    We note that the magnonic analog of quantum Hall effect has been investigated on the square lattice in Ref. \cite{mei}. The authors have elaborated on the non-relativistic low-temperature magnonic Hall conductance due to the lowest Landau level.  Basically, when the temperature is very small compare to the energy level spacing, only the lowest Landau level contributes to the magnon Hall conductance and the Bose occupation function $n_B$ can be approximated as a constant for a nearly flat band. Consequently, the magnon Hall conductance $G_{xy}$ in the non-relativistic system is quantized as \cite{mei} $G_{xy}= \frac{(g\mu_B)^2}{h}n_B(E_0^*)\cdot N$, where $N \in \mathbb{Z}$ and $E_0^*$ is the energy of a nearly flat band.  A similar argument can be applied to the relativistic Dirac magnon, but before we dive into more discussion let us first understand the Landau quantization of the Dirac magnons.  

 \subsection{Quantum field theory of  Dirac magnon in the  weak electric field regime}
 
In quantum field theory, it is well-known that neutral particles with magnetic dipole moment can couple to an external electromagnetic field.  For charge-neutral magnons in the 2D insulating honeycomb ferromagnet, the low-energy excitation near the band touching point (say $-\text{\bf K}$) can be described by the 2D Dirac equation.  Upon exposure to a weak electromagnetic field, the system is governed  by the Dirac-Pauli Lagrangian  \cite{bjo, paul},

\begin{align}
 \mathcal L=\bar\Psi\Big[-v_0\gamma^0 + i\gamma^0\partial_\tau+iv_D\gamma^i\partial_i-\frac{v_D\mu_m}{2}\sigma^{\mu\nu} F_{\mu\nu}\Big]\Psi,
 \label{Lag}
\end{align}
where $v_0 = J_0/2$ accounts for the finite energy Dirac point, whereas $v_D = J_0/4$ is the group velocity near the Dirac point.  The  two-component wave function is given by $\Psi = \big( u_A(\vec r, \tau),u_B(\vec r, \tau) \big)$, and $\bar\Psi=\Psi^\dg\gamma^0$.  The electromagnetic field tensor is $F_{\mu\nu}$   and  $\sigma^{\mu\nu}=\frac{i}{2}[\gamma^\mu,\gamma^\nu]=i\gamma^\mu\gamma^\nu,~ (\mu\neq \nu)$ with $\gamma^{\mu}=(\gamma^0,\gamma^i)$.  The corresponding Hamiltonian is derived in Appendix~A.  We consider one of the two representations of gamma matrices in (2+1) dimensions:
\begin{align}
\gamma^0=\sigma_z,~\gamma^1= i\sigma_x,~\gamma^2= -i\sigma_y,
\end{align}
where $\sigma_i$ are Pauli matrices. They  satisfy the following relations: 
\begin{align} 
 \lbrace \gamma^\mu,\gamma^\nu\rbrace=2g^{\mu\nu},
 \end{align}  
where $g^{\mu\nu}=\text{diag}(1,-1,-1)$ is the Minkowski metric in (2+1) dimensions.

\begin{figure*}
\centering
\includegraphics[width=1\linewidth]{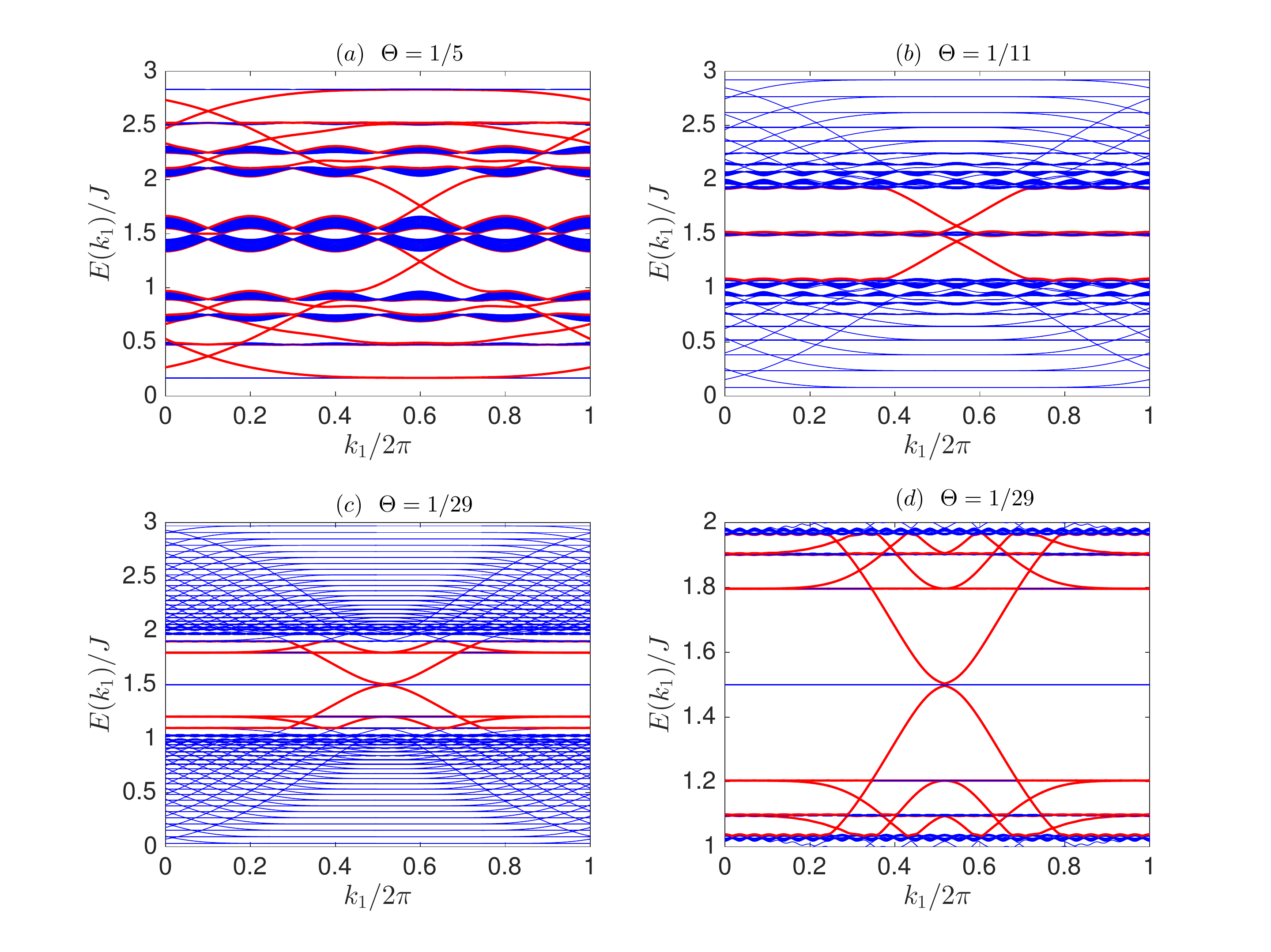}
\caption{Color online. The magnonic energy spectra of the honeycomb-lattice ferromagnet with zigzag edge for (a) $\Theta=1/5$, (b) $\Theta=1/11$,  (c) $\Theta=1/29$, (d) zoom in of   (c).}
\label{edge}
\end{figure*}

 \begin{figure}
\centering
\includegraphics[width=1\linewidth]{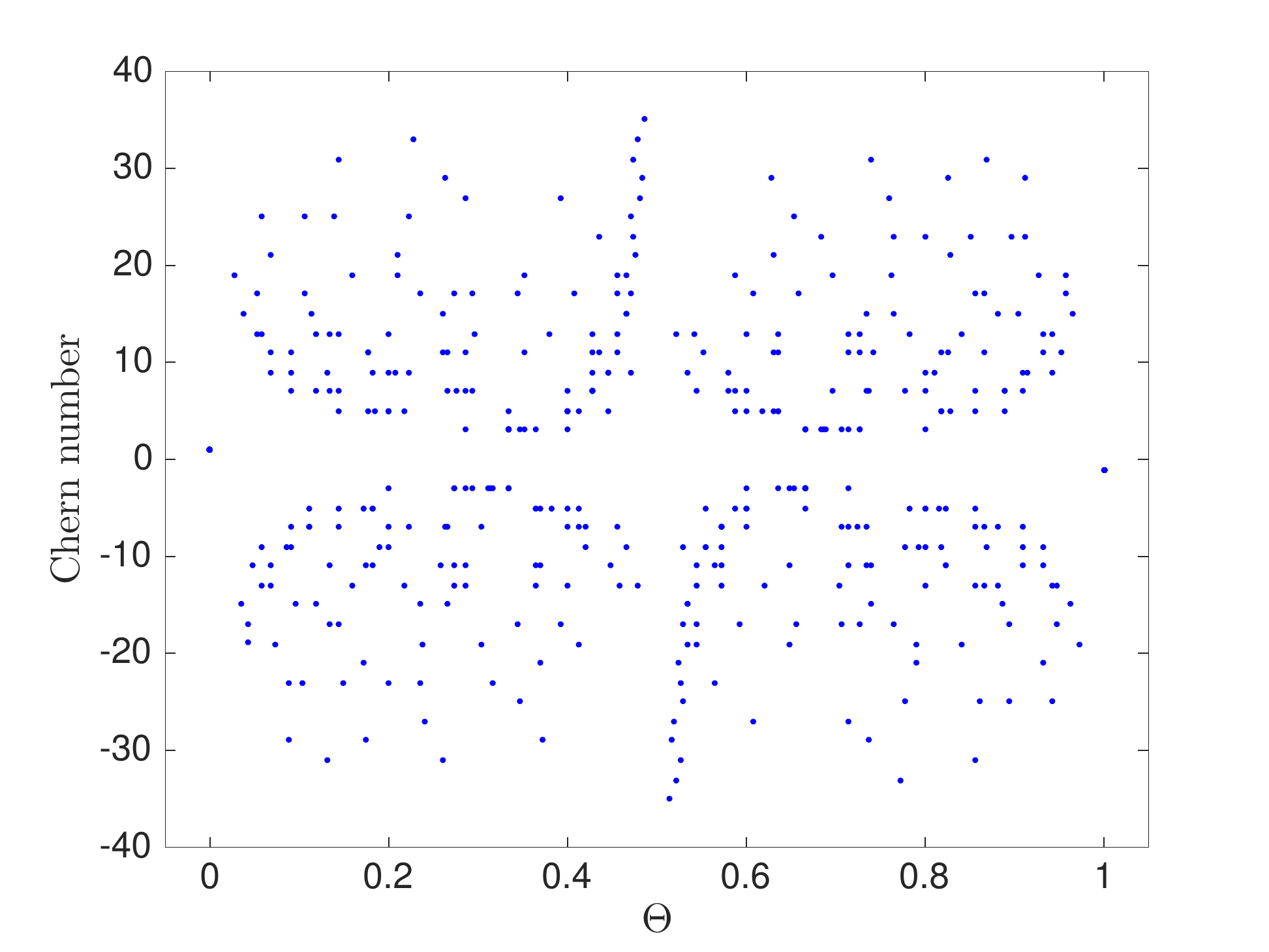}
\caption{Color online. Magnonic Chern number distribution for the ground state of the Hofstadter butterfly ({\it i.e.}, the lowest Hofstadter spectrum) as a function of $\Theta = \frac{p}{q}$ for $q<41$.}
\label{Chern_Butterfly}
\end{figure}

The Lagrangian formalism applies to a general electromagnetic field. Let us consider an electromagnetic field with only  spatially-varying electric field vector $\vec{E}(\vec r)$. In this case $\frac{1}{2}\sigma^{\mu\nu} F_{\mu\nu} = i\gamma^0\gamma^i E_i$.  The corresponding equation of motion takes the form

\begin{align}
\big[-v_0\gamma^0 +i\gamma^0\partial_\tau + iv_D\gamma^i\big(\partial_i -\mu_m\gamma^0 E_i\big) \big]\Psi = 0.
\label{EOM}
\end{align}
To be specific, let us  consider   $\vec E(\vec r) = \lambda (0, -y, 0)$. Then the general solution to Eq.~\ref{EOM} can be written as $\Psi(\vec r, \tau) = e^{-i\varepsilon \tau + ik x}\psi(\xi)$, where $\xi = y/l_{\lambda} + kl_{\lambda}~\text{sign}(\mu_m\lambda)$ is a dimensionless coordinate replacing $y$, and $l_{\lambda} = 1/\sqrt{|\mu_m\lambda|}$ is the electric length. The equation of motion \eqref{EOM} for the two-component spinor $\psi(\xi)$ now takes the form
\begin{widetext}
\begin{align}
\begin{pmatrix}
\varepsilon-v_0  & \frac{iv_D}{l_{\lambda}}\big[\frac{d}{d\xi}-\xi\text{sign}(\mu_m\lambda)\big]\\
-\frac{iv_D}{l_{\lambda}}\big[\frac{d}{d\xi}+\xi\text{sign}(\mu_m\lambda)\big] & -\varepsilon + v_0 
\end{pmatrix}{u_A\choose u_B}= 0.
\label{EOM1}
\end{align}
\end{widetext}
The resemblance of this equation  to that of relativistic charged electron in a perpendicular magnetic field \cite{vp, mir} is evident. 
For  concreteness, we assume that $\mu_m\lambda>0$, then the solution to the eigenvalue equation yields \cite{ak,vp,mir}
\begin{align}
&\varepsilon_0 = v_0\\&
\varepsilon_n =v_0 \pm v_D\sqrt{2|\mu_m\lambda|n}, \quad n = 1,2,\cdots
\end{align}

These are the magnonic Landau levels for the 2D insulating honeycomb ferromagnet in the low electric field regime. This result was obtained conventionally in the appendix of Ref.~\cite{mei}.  Here, we have provided an intuitive derivation from quantum field theory.  Since there is no spin degeneracy in the present system, the Landau levels have only valley degeneracy at the $\pm\bf{K}$ point.  Evidently, the honeycomb magnonic Hofstadter-Landau level spectra in Fig.~\ref{HCB_Butterfly}  is similar to  the electronic counterpart \cite{ram,yas, gold,hat}. The only difference  in the fractal structure is that the Dirac magnon and the zeroth Landau level $\varepsilon_0$ occur at nonzero energy as expected in the bosonic systems.  However, the topological invariant of the system is unaffected by the finite Dirac magnon energy, which is only a reference point to the zero energy mode in graphene.  In other words, the finite zeroth Landau level in the current system is synonymous with the zero energy mode in graphene that leads to an unconventional quantum Hall effect \cite{gus,nov}.  This similarity is evident since the finite zeroth Landau level $\varepsilon_0$ also has twice smaller degeneracy than the levels with $n > 0$, as it is shared between the bands above and below the Dirac magnon energy.  This implies that magnonic Hall conductance in the relativistic Dirac magnon is $G_{xy}= \pm \frac{(g\mu_B)^2}{h}n_B(E_0^*)\cdot C_H$, where $C_H = 2N +1$ is the Chern number obtained numerically using the method described in Ref.~\cite{fuk, hat}.

\subsection{The bulk-edge correspondence}

To get a lucid picture of the discrete odd Chern numbers in the 2D insulating honeycomb ferromagnet, we now present the bulk-edge correspondence of the system. The basic idea is  that the Chern number $C_H = 2N +1$ is related to the topological number $I_H$ for edge states which corresponds to the number of intersections on the Riemann surface $\Sigma_{2q-1}$ with $2q-1$ energy gaps \cite{hat}. In order to investigate this correspondence, we have solved for the edge state modes using a cylindrical geometry periodic along the $y$ direction and infinite along the $x$ direction. The result is depicted in Fig.~\ref{edge} for various values of $\Theta = p/q$. 

 In Fig.~\ref{edge}(a), we have shown the ten magnonic  bands corresponding to strong electric field for $\Theta = 1/5$. In this case, the middle bands touch  and they are accompanied by a dispersionless edge state mode at the Dirac magnon point. Therefore, they do not contribute to the topological number. The other bands have disperse edge states traversing the gap. The topological number is given by the number of edge state modes crossing at each gap, which is given by $I_H= 1, 2, 3, -1, +1, -3, -2, -1$. 

In Fig.~\ref{edge}(c), we have shown the magnonic bands corresponding to weak electric field for $\Theta = 1/29$. The zoom-in figure is depicted in Fig.~\ref{edge}(d). This case is more realistic.  We can see that the magnonic bands are nearly flat and look more like the magnonic Landau levels, with edge states traversing the gap.  In the vicinity of the Dirac magnon, we can see that the topological number is given by $I_H= 1, 3, 5, \cdots$ which corresponds to the Chern numbers at weak electric field.

In fact, the ground state of the magnonic Hofstadter spectrum (i.e. the lowest Hofstadter spectrum in Fig.~\ref{HCB_Butterfly}) also exhibits odd Chern numbers.  Following Refs.~\cite{wac} and \cite{gold}, we have numerically \cite{fuk} obtained the ground state odd Chern number of the lowest Hofstadter spectrum as shown in Fig.~\ref{Chern_Butterfly} for $q < 41$.  Indeed, the distribution trend of the Chern number is consistent with that of electronic honeycomb lattice (see Fig.~7 in Ref.~\cite{wac}). However, it is important to note that the lowest Hofstadter spectrum in the present case corresponds to the band close to the Goldstone zero energy mode.   The consistency of the present results with electronic honeycomb lattice shows that the concept of  Chern number of topological bands is completely independent of the statistical nature of the quasiparticle excitations.

 

\begin{figure*}
\centering
\includegraphics[width=1\linewidth]{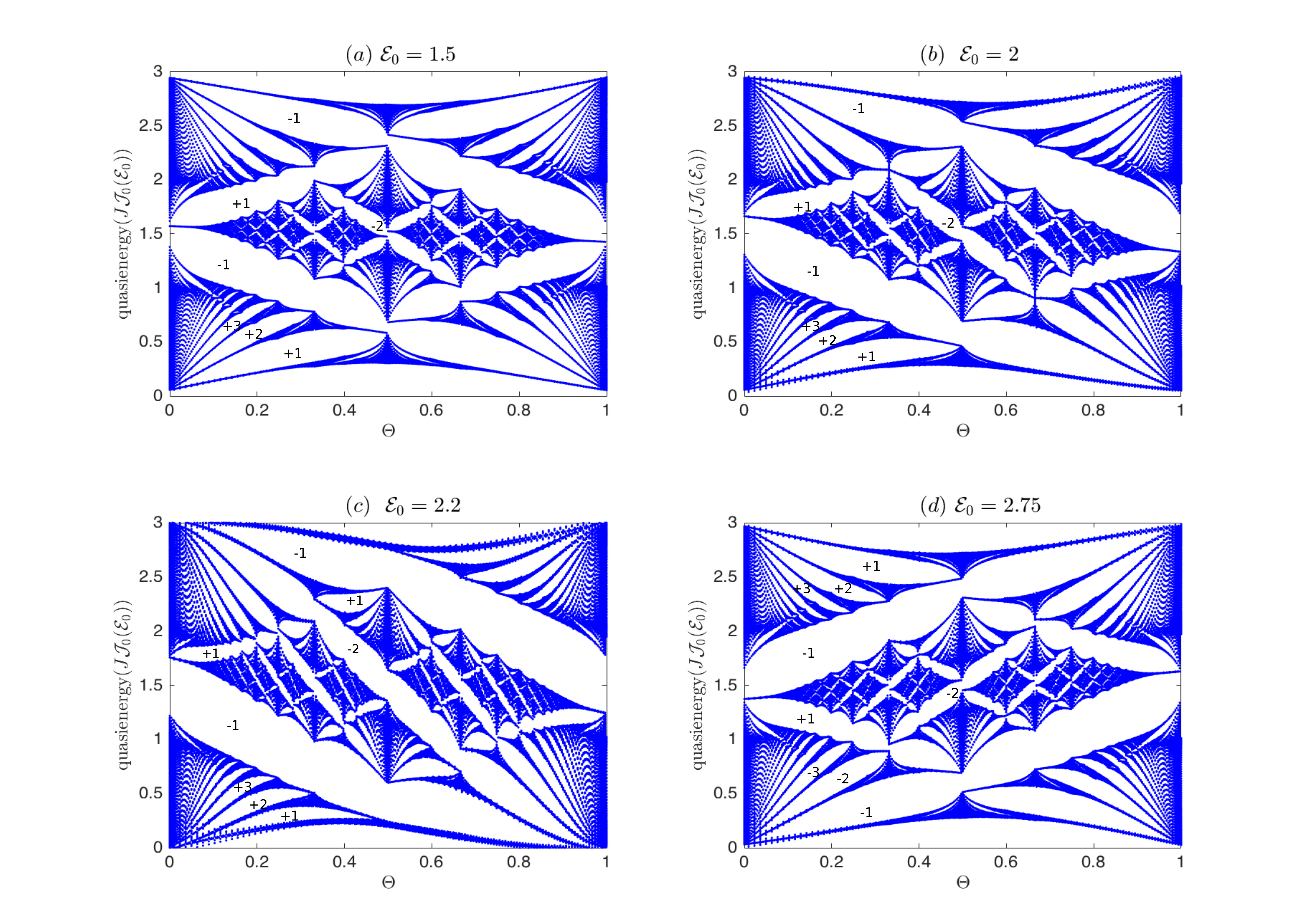}
\caption{Color online. Topological phase transition in magnonic Floquet  Hofstadter spectra  for varying  amplitude  $\mathcal E_0$ in units of $g\mu_Ba/\hbar c^2$ at high frequency $\omega/J=6$.}
\label{Floquet_butterfly_HF}
\end{figure*}

\begin{figure}
\centering
\includegraphics[width=1\linewidth]{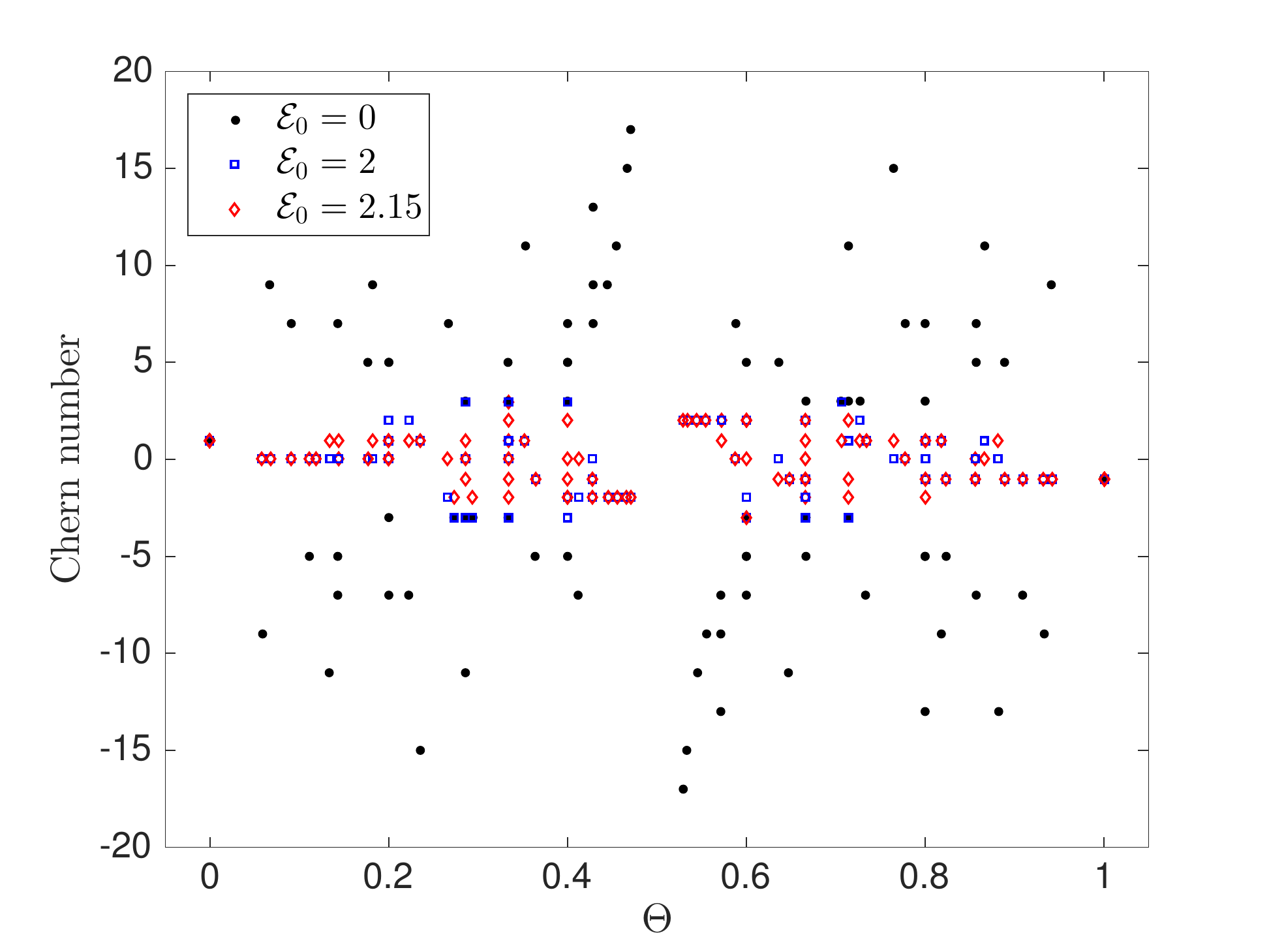}
\caption{Color online.  Magnonic Floquet Chern number distribution for the ground state of the Hofstadter butterfly ({\it i.e.}, the lowest Hofstadter spectrum) as a function of $\Theta = \frac{p}{q}$ for $q\leq 21$ and $\omega/J= 6$.}
\label{Floquet_ChernN}
\end{figure}

\section{Floquet  magnonic Hofstadter spectrum}

\subsection{ Time-dependent Aharanov Casher phase}

Having studied the magnonic Hofstadter spectrum in the presence of a time-independent  spatially-varying electric field, we now investigate the interaction of the magnonic Hofstadter spectrum  with light. We will consider light with dominant oscillating electric field component ${\vec E}(\vec r, \tau)$, irradiated perpendicular to the 2D insulating honeycomb  ferromagnet. The resulting effect is that magnon hopping on the insulating magnet  will accumulate the time-dependent Aharanov Casher phase
\begin{align}
\theta_{mn}(\tau) = \mu_m\int_{\vec R_{mn}}^{\vec R_{mn}^\prime}  \mathscr {\vec E}(\vec r, \tau) \cdot d{\vec l},
\label{acp}
\end{align}
where  $\mathscr {\vec E}(\vec r, \tau)={\vec E}(\vec r, \tau) \times\hat z$. The oscillating  electric field is defined as 
\bea 
{\vec E}(\vec r,\tau)=-{\vec \nabla}\phi(\vec r,\tau)-\partial_\tau {\vec A}(\vec r,\tau),
\label{Efield}
\eea 
where $\phi(\vec r,\tau)$ is the electromagnetic scalar potential and   ${\vec A}(\vec r,\tau)$ is the vector potential.  We assume a periodic oscillating  electric field, {\it i.e.}  ${\vec E}(\vec r,\tau)={\vec E}(\vec r,\tau +T)$  with period  $T=2\pi/\omega$.  In the case of very weak spatial variation,  the oscillating electric field ${\vec E}(\tau)=-\partial_\tau \vec{A}(\tau)$ is the driving force in both electronic \cite{foot5, foot3} and magnonic \cite{owe0, owe1, owe2, sat} Floquet topological systems. Hence, the Floquet states arising from both cases are  direct analogs.  The time-periodic magnon phase enters the Hamiltonian through the Peierls substitution as in Eq.~\ref{phase}. Consequently, the spin or the bosonic Hamiltonian becomes  time-dependent, and thus can be studied by the Floquet-Bloch theory in the same manner as driven  electronic charged systems \cite{foot5, foot3}.

\begin{figure*}
\centering
\includegraphics[width=1\linewidth]{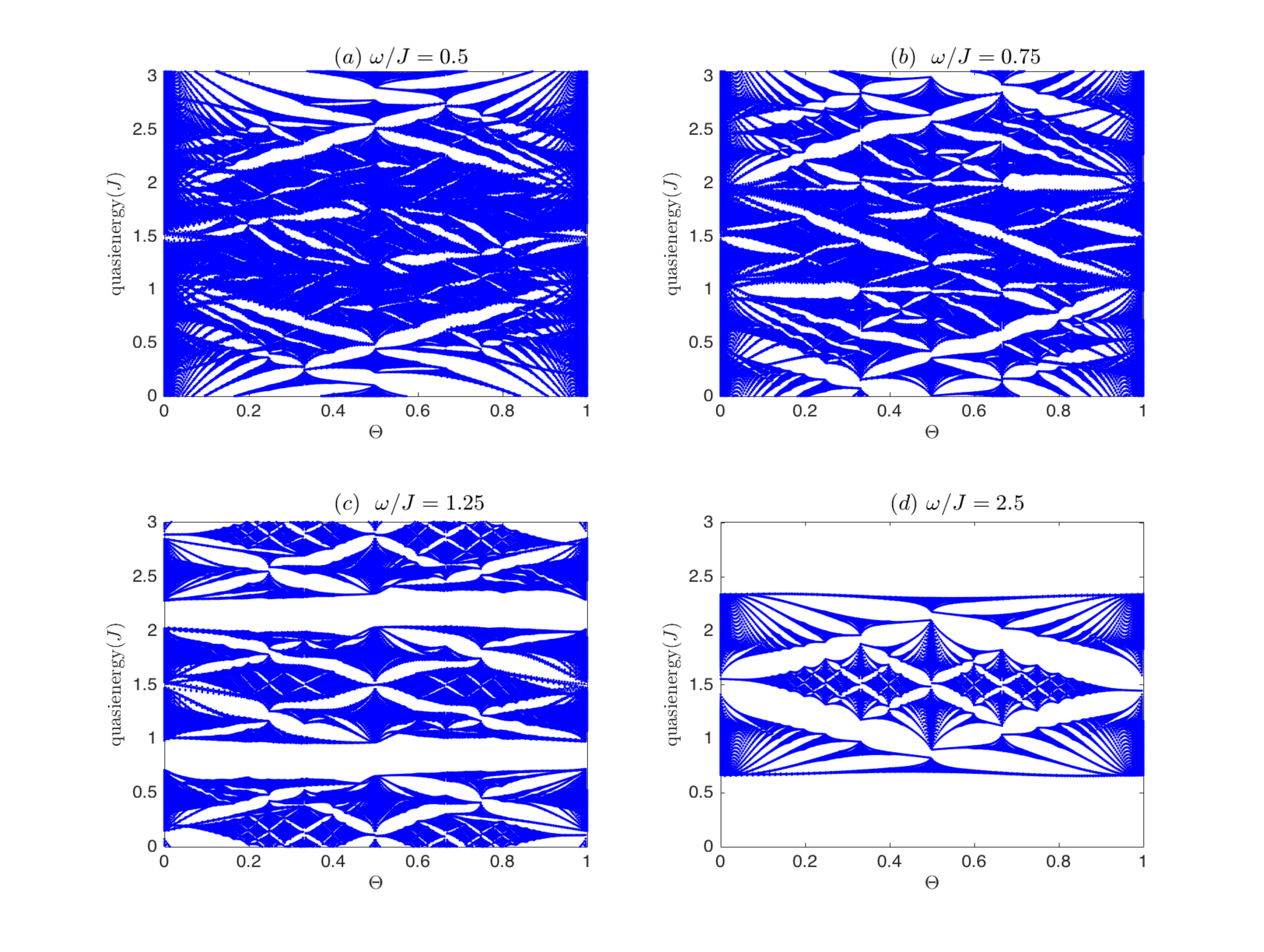}
\caption{Color online. Magnonic Floquet  Hofstadter spectra  for varying frequency at fixed amplitude $\mathcal E_0=1$.}
\label{Floquet_butterfly}
\end{figure*} 

 We assume that the oscillating space- and time-dependent electric field is derivable from the scalar and vector potentials. Hence, we write them as
 \begin{align}
 &\phi(\vec r,\tau)  = \phi(\vec r)= \frac{\lambda}{2}y^2,\\&\vec A(\vec r,\tau) = \vec A(\tau) = A_0\big[\sin(\omega\tau), \cos(\omega\tau),0\big],
 \end{align}
where $A_{0}= E_{0}/\omega$ is the strength of the time-dependent circularly-polarized vector potential.  Recall that $\mathscr {\vec E}(\vec r, \tau)={\vec E}(\vec r, \tau) \times\hat z$, therefore using Eq.~\ref{Efield} we obtain a separable space- and time-dependent oscillating electric field
\begin{align}
&\mathscr {\vec E}(\vec r, \tau) =\big[E_0\sin(\omega \tau)- \lambda y,   E_0\cos(\omega \tau) , 0\big],
\label{osci}
 \end{align}
 where $E_0$ is the strength of the time-dependent circularly-polarized  electric field. It is crucial to point out that the present formalism is different from that of driven electronic system.  In the latter, the Floquet  Hofstadter spectrum  is generated by independently applying an oscillating electric field and a perpendicular magnetic field  \cite{wac, wac1, wac2}. In the magnonic system, however, the Floquet  Hofstadter spectrum is a consequence of the separable oscillating (space and time-dependent) electric field as given by Eq.~\ref{osci}. In this case, the role of a perpendicular magnetic field is played by the harmonic electromagnetic scalar potential, which is also directly encoded in the  electric field via Eq.~\ref{Efield}.

The resulting time-dependent Hamiltonian is given by
\begin{align}
\mathcal H(\tau)&=J_0S\sum_{mn} \big(a_{m,n}^\dagger a_{m,n}+b_{m,n}^\dagger b_{m,n}\big)\nonumber\\&-JS\sum_{ mn}\Big[a_{m,n}^\dagger\big(e^{i\pi\frac{\Theta}{\Theta_0}}e^{i\mu_m\mathscr{\vec E}(\tau)\cdot \vec \delta_3}b_{m,n}\nonumber\\&+e^{-i\pi\frac{\Theta}{\Theta_0}}e^{i\mu_m\mathscr{\vec E}(\tau)\cdot \vec \delta_1}b_{m-1,n}\nonumber\\&+e^{i\mu_m\mathscr{\vec E}(\tau)\cdot \vec \delta_2}b_{m,n-1}\big) + \text{H.c.}\Big],
\label{timedep}
\end{align} 
where $\mathscr{\vec E}(\tau)$ is the time-dependent part of the electric field $\mathscr{\vec E}(\vec r, \tau)$. As we did before, the time-dependent Hamiltonian in Eq.~\ref{timedep}  can be written as a $2q\times 2q$ matrix in momentum space.


\subsection{Magnonic Floquet Hofstadter Hamiltonian}
We will now apply the Floquet-Bloch formalism to the problem. The  time-periodic momentum space Hamiltonian $\mathcal H(\vec{k},\tau)$ can be expanded  as
\begin{align}
\mathcal H(\vec{k},\tau)=\sum_{n=-\infty}^{\infty} e^{in\omega \tau}\mathcal H_n(\vec{k}),
\end{align}
where the Fourier components  are given by

\begin{align}
\mathcal H_n(\vec{k})=\frac{1}{T}\int_{0}^T e^{-in\omega \tau}\mathcal H(\vec{k}, \tau) d\tau=\mathcal H_{-n}^\dg(\vec{k}).
\label{fcom}
\end{align} 
The corresponding  eigenvectors can be written as $\ket{\psi_\alpha(\vec{k}, \tau)}=e^{-i \epsilon_\alpha(\vec{k}) \tau}\ket{\chi_\alpha(\vec{k}, \tau)},$
where  $\ket{\chi_\alpha(\vec{k}, \tau)}=\ket{\chi_\alpha(\vec{k}, \tau+T)}=\sum_{n} e^{in\omega \tau}\ket{\chi_{\alpha}^n(\vec{k})}$ is the time-periodic Floquet-Bloch wave function of magnons and $\epsilon_\alpha(\vec{k})$ are the magnon quasi-energies. We define the Floquet operator  as $\mathcal H^F(\vec{k},\tau)=\mathcal H(\vec{k},\tau)-i\partial_\tau$.  The corresponding eigenvalue equation is of the form 
\begin{align}
\sum_m \big[\mathcal H_{n-m}(\vec{ k}) + m\omega\delta_{n,m}\big]\chi_{\alpha}^m(\vec{k})= \epsilon_\alpha(\vec{k})\chi_{\alpha}^n(\vec{k}).
\end{align}
Next, we calculate the Fourier components of the $2q\times 2q$  Hamiltonian $\mathcal H_{q,\ell}(\vec{k},\tau)$ with $\ell = m-n$.  We obtain

\begin{align}
\mathcal H_{q,\ell}(\vec k)&=J_0S\delta_{\ell,0}\mathbb{I}_{2q\times 2q}-JS
\begin{pmatrix}
0&\mathcal{F}_\ell(\bo)\\
\mathcal{F}_{-\ell}^\dg(\bo)&0\\
\end{pmatrix},
\label{fham1}
\end{align}
where $\mathcal{F}_\ell({\vec k})$ is a $q\times q$ matrix given by

\begin{align}
&\mathcal{F}_\ell({\vec k}) = \big[\mathcal J_\ell(\mathcal E_0)e^{i\pi\ell/2}\mathcal{U} +\mathcal J_{-\ell}(\mathcal E_0)e^{-i\pi\ell/6}e^{ik_1}\mathcal{U}^\dg\big] \\&\nonumber+ \mathcal J_\ell(\mathcal E_0)e^{i\pi\ell/6} \mathcal{V}.
\end{align}
Here  $\mathcal{U} = U$ with $\xi_r = e^{ir\pi\frac{\Theta}{\Theta_0}}$,  $\mathcal{V} = V$, and  $\mathcal J_\ell(x)$ is the Bessel function of order  $\ell =1,0,-1$.  It is easy to check that Eq.~\ref{fham1} reduces to Eq.~\ref{free} in the  absence of the oscillating electric field. 

Another distinguishing feature of the  magnonic Floquet formalism is that the light intensity  is characterized by the dimensionless quantity
\begin{align}
\mathcal E_0 =\frac{g\mu_B E_0 a}{\hbar c^2}.
\label{dims}
\end{align}
We can see that  there is  no frequency denominator in Eq.~\ref{dims}, as opposed to the case of electronic Floquet formalism \cite{foot3, foot5}.

\subsection{ Topological phase transitions in the magnonic Floquet Hofstadter spectrum}

Now, we will study the deformation of the fractal structure due to light irradiation and the associated topological phase transitions. For this purpose, let us first consider the high-frequency limit, when the frequency of the light  $\omega$ is much greater than the magnon band-width $\sim 3J$.  In this regime, the Floquet sidebands are completely decoupled and the system can be described by an effective time-independent  Hamiltonian \cite{tp6,tp7,tp8} by  expanding perturbatively in $1/\omega$. The resulting effective  Hamiltonian is given by
\begin{align}
\mathcal H^{\text {eff}}_q(\vec k)&=\mathcal H_{q,0}(\vec k)-\frac{1}{\omega}\big[\mathcal H_{q,-1}(\vec k), \mathcal H_{q, 1}(\vec k)\big],
\label{effHam}
\end{align}
where $\mathcal H_{q,0}(\vec k)$ is the zeroth-order Hamiltonian and $\mathcal H_{q, \pm 1}(\vec k)$ are the single photon dressed Hamiltonians.

 In Fig.~\ref{Floquet_butterfly_HF} we have plotted the magnonic Floquet  Hofstadter spectra in the high-frequency regime $\omega = 6J$ as a function of $\Theta$, for several values of the amplitude $\mathcal E_0$ in units of $g\mu_Ba/\hbar c^2$.   We have rescaled  the magnon quasienergy $\epsilon_{\alpha}(\bo)$ by the effective coupling $J\mathcal{J}_0(\mathcal E_0)$. We can see that the symmetry of the static  magnonic  Hofstadter spectra in Fig.~\ref{HCB_Butterfly} is completely broken due to explicit time-reversal symmetry breaking by circularly-polarized light. It is crucial to first understand the effects of circularly-polarized light in the absence of the flux $\Theta =0$, i.e. no static electric field. In this case the system realizes a magnonic Floquet topological insulator \cite{owe0}, which stems from a photo-induced  synthetic scalar spin chirality or a staggered Dzyaloshinskii-Moriya (DM) interaction \cite{dm,dm2} along the $z$-axis. The photo-induced DM interaction breaks the time-reversal symmetry of the relativistic Dirac magnons and induces a topological band gap with Chern number $ C_F = \pm \text{sign}(D_F) = \pm 1$ for the top and bottom magnon bands respectively, where $D_F = \frac{\sqrt{3}J^2\mathcal J_{1}(\mathcal E_0)^2}{\omega} \approx \frac{\sqrt{3}}{2} \frac{J^2 \mathcal E_0^2}{\omega}$. 
 
 As we can see from Fig.~\ref{Floquet_butterfly_HF} (a) -- (c), the gap at the Dirac magnon energy $\epsilon_D \sim 1.5$ persists as $\Theta$ is continuously tuned from $0$ to $1$, but it is now connected with a large gap below $\epsilon_D$ with a Chern number of $-1$. As the amplitude of light is increased, there is a signal of a topological phase transition as we approach the first zero of $\mathcal{J}_0(\mathcal E_0)$ at $\mathcal E_0\sim 2.4$. The signal of this topological phase transition is clearly evident by the change in the fractal structure at $\mathcal E_0= 2.2$ as shown in Fig.~\ref{Floquet_butterfly_HF}(c).  It is accompanied by a gap closing  point as we approach the first zero of $\mathcal{J}_0(\mathcal E_0)$. As shown in Fig.~\ref{Floquet_butterfly_HF}(d)  the gap reopens for $\mathcal E_0= 2.75$. However, we can see that the gap at the Dirac energy $\epsilon_D \sim 1.5$  for $\Theta \neq 0$ is now connected with a large gap above $\epsilon_D$ with a Chern number of $+1$. In other words, there is a band inversion at the topological phase transition.  Apart from the topological phase transition associated with varying  $\mathcal E_0$, there are also other topological phase transitions at fixed $\mathcal E_0$ as shown in Fig.~\ref{Floquet_butterfly_HF}. For instance in the vicinity of $\epsilon_D$, non-trivial gap opens at the fluxes $\Theta = 1/2, 1/3, 1/5$, and the Chern number in this case is equal to the denominator of the rational flux. In fact, there is a unique ground state in the Floquet Hofstadter spectrum in the high-frequency regime \cite{wac}. Therefore, we numerically calculate the ground state Chern number of the lowest Floquet Hofstadter spectrum for several values of the amplitudes $\mathcal E_0$ as a function of $\Theta =p/q$, using the discretize BZ method \cite{fuk}.     In Fig.~\eqref{Floquet_ChernN} we show the results  for $q\leq 21$. Indeed, the distribution trend of the Chern number changes in the presence of light, with similar trend to that of driven electronic honeycomb lattice \cite{wac}.

In Fig.~\ref{Floquet_butterfly} we have shown the plots of the magnonic Floquet  Hofstadter spectra for varying low-frequencies less than the magnon band-width $\sim 3J$ at fixed amplitude $\mathcal E_0 = 1$ in units of $g\mu_Ba/\hbar c^2$. At very low-frequencies $\omega/J =0.5$ and $\omega/J =0.75$ as shown in Fig.~\ref{Floquet_butterfly}(a) and Fig.~\ref{Floquet_butterfly}(b) respectively, the magnonic Floquet Hofstadter mutibands begin to interact with each other. In the vicinity of the band crossings, we can see that there are still topological gaps, but the Chern number is difficult to calculate numerically since there is no unique ground state. As the frequency is slightly increased  (see Fig.~\ref{Floquet_butterfly}(c) with $\omega/J =1.25$), the magnonic Floquet Hofstadter mutibands start to separate. Eventually,  they are completely decoupled  for frequency comparable to the band-width  $\sim 3J$ as shown in Fig.~\ref{Floquet_butterfly}(d) for $\omega/J =2.5$.

\section{Conclusion}
We have presented an exposition of magnonic Floquet Hofstadter butterfly in the 2D insulating honeycomb ferromagnet. We formulated the  theory based on the Aharonov-Casher phase acquired by hopping charge-neutral magnons in the presence of an oscillating space- and time-dependent electric field.  We presented results for both static and periodically  driven magnonic Hofstadter butterfly. We also studied the rich topological phase transition associated with different gap openings and band inversion in the magnonic Floquet Hofstadter spectra, and computed the ground state odd Chern number distribution of the lowest magnonic Hofstadter spectrum.

In addition to the magnonic Hall conductance, which is due to a magnetic field gradient \cite{fuji}, we believe that the  magnonic Floquet Hofstadter spectrum will also exhibit the thermal Hall effect \cite{thm1, thm2, thm3,thm4} by applying a temperature gradient on the driven 2D insulating magnet. We believe that the current results will pave the way for manipulating  the intrinsic property of 2D insulating honeycomb magnets such as CrX$_3$ (X = Br, Cl, I), using circularly-polarized light. Furthermore, the predicted results  are pertinent to new experiments and will remarkably impact future research in this field with potential practical applications to photo-magnonics \cite{benj}, magnon spintronics \cite{chu, benja},  and ultrafast optical control of magnetic spin currents \cite{ment, tak4, tak4a,walo}. 

\section*{ACKNOWLEDGEMNTS}

I would like to thank  M. Wackerl for useful discussion. Research at Perimeter Institute is supported by the Government of Canada through Industry Canada and by the Province of Ontario through the Ministry of Research
and Innovation. 

\section*{APPENDIX A: DIRAC MAGNON HAMILTONIAN IN THE PRESENCE OF AN ELECTRIC FIELD}
\label{Appen}
 In this appendix, we derive the Dirac Hamiltonian in the presence of a general electric field $\vec{E}(\vec r, \tau)$. This will justify how the Aharonov-Casher phase enters the Hamiltonian  as a minimal coupling in (2+1) dimensions. Starting from the Lagrangian density in Eq.~\ref{Lag}, the Hamiltonian is given by
 \begin{align}
 H = \int d^2 x ~\big [ \pi(x)\dot{\Psi}(x) - \mathcal L \big ] \equiv \int d^2 x ~ \Psi^\dg \mathcal H_D \Psi,
 \end{align}
 where $\pi(x) = \frac{\partial \mathcal L}{\partial{\dot{\Psi}(x)}}$ 
is the generalized momentum. The Dirac magnon Hamiltonian is given by
\begin{align}
 \mathcal H_D=v_0 - iv_D\vec{\alpha}\cdot \vec{\nabla}+\gamma^0\frac{v_D\mu_m}{2}\sigma^{\mu\nu} F_{\mu\nu},
 \label{hamil}
\end{align}
where $\vec{\alpha}=\gamma^0\vec{\gamma}$. 

For an electromagnetic field with only an electric field vector $\vec{E}(\vec r, \tau)$, we have $\frac{1}{2}\sigma^{\mu\nu} F_{\mu\nu} = i\vec{\alpha} \cdot \vec E(\vec r, \tau)$. Under the unitary transformation $\sigma_x \leftrightarrow \sigma_y$, the  Dirac magnon Hamiltonian can be written as

\begin{align}
\mathcal H_D=v_0 +v_D\vec{\sigma}\cdot\big[-i\vec{\nabla} + \mu_m\vec{E}(\vec r, \tau)\times \hat z\big].
\end{align}
We can see that the Aharonov-Casher phase enters the relativistic Dirac magnon Hamiltonian  as a minimal coupling in (2+1) dimensions.

\end{document}